\documentstyle[12pt,epsf]{article}

\large
\newcommand{\be}{\begin{eqnarray}}
\newcommand{\ee}{\end{eqnarray}}
\newcommand\del{\partial}

\newcommand\eg {{\it e.g. }}

\newcommand\half{\frac 1 2 }
\newcommand\noi {\noindent}

\begin{document}
\setlength{\baselineskip}{21pt}
\pagestyle{empty}
\vfill
\eject
\begin{flushright} USITP-98-03 \\
March 1998
\end{flushright}
\mbox{}
\vskip 1.5cm
\centerline{\Large\bf Nonperturbative SUSY Correlators at Finite Temperature}
\vskip 1.2 cm
\begin{center}
{\bf J. Grundberg} \\
Department of Mathematics and Physics \\
M\"{a}lardalens H\"{o}gskola \\
Box 833, S-721 23 V\"{a}ster{\aa}s, Sweden 
\vskip 3mm
{\bf J. Wirstam}  \\
Institute of Theoretical Physics\\
University of Stockholm \\
Box 6730, S-113 85 Stockholm, Sweden \\
\end{center}

\vskip 1cm\noi
\centerline{\bf ABSTRACT}
\vskip .5cm
We calculate finite temperature effects on a 
correlation function in the two dimensional supersymmetric
nonlinear O(3) sigma model. The correlation function violates chiral symmetry
and at zero temperature it has been shown to be a constant, which gives rise
to a double--valued condensate. Within the bilinear approximation we find
an exact result in a one--instanton background at finite temperature. In contrast
to the result at zero temperature we find that the correlation function decays
exponentially at large distances. 

\vfill
\eject
\pagestyle{plain}

\newcommand \streck {\overline} 
\newcommand \zbar {\bar{z}}
\newcommand \psibar {\overline\psi}
\newcommand \thetaf [4] {\Theta\left[\begin{array}{c}#1 \\ #2
\end{array}\right] (#3 , #4) }
\newcommand \Dsl {D\!\!\!\!/}
\newcommand \dsl {{\partial}\!\!\!\!/}
\setlength{\baselineskip}{23pt}

\noindent{\bf 1. Introduction }
\renewcommand{\theequation}{1.\arabic{equation}}
\setcounter{equation}{0}
\vskip 5mm

During the last few years there has been a lot of progress in understanding the dynamics of supersymmetric (SUSY) theories.
Although the main achievements have been obtained in $N=2$ supersymmetry, where \eg the low energy effective action can be 
calculated exactly \cite{seiwitt}, there also exist some remarkable results in $N=1$. Among these are the dual descriptions
of supersymmetric QCD, originally proposed by Seiberg \cite{n1seiberg}.

Related to this last issue is the calculation of certain instanton induced correlation functions in $N=1$ SUSY Yang--Mills theory
\cite{Novikovetal, amatietal}. The action for this theory possesses a global R--symmetry that is conserved
classically, but broken by quantum effects. Even though the symmetry is anomalous, there still 
remains a nonanomalous discrete subgroup. Some specific correlation functions that violate chiral symmetry, and therefore
vanish to all orders in perturbation theory, can be calculated in an instanton background for small distances
 and are found to have a constant value, independent of the actual distance. 
The reason for this nonvanishing value is of course that the instantons change the chirality by a definite value. 
Furthermore, by using supersymmetric arguments it was shown that this result actually is {\em exact}, for all distances.
By the cluster decomposition this implies a nonvanishing value of the gluino condensate, $\langle \lambda \lambda \rangle $, and
the nonvanishing value spontaneously breaks the discrete symmetry down to $Z_2$. As has recently been noted, this condensate
gives rise to domain wall solutions \cite{domains}.

It is a well known feature that if a symmetry is spontaneously broken at zero temperature, it is
often restored above some, possibly high, temperature. This is believed to be the case \eg for the Higgs mechanism in the 
Electroweak Standard Model \cite{linde} and for the spontaneously broken chiral
symmetry in QCD \cite{bernard}. In view of this, it is clearly interesting to study also the finite temperature effects on 
the spontaneously broken discrete
symmetries in supersymmetry, especially when one considers the solid results that already exist at $T=0$.

Since the exact results at zero temperature depend crucially on specific properties of the supersymmetric theory, and since
these relations in general are not valid at finite temperature, one should not expect the $T>0$ results to be constrained 
in the same precise manner as at $T=0$. However, the underlying structure of these nontrivial theories may still be simple enough in order
to obtain interesting results.

In the SUSY SU(2) Yang--Mills theory at finite temperature, some preliminary results were recently obtained \cite{jimojag}, but no definite, conclusive
arguments could be given. In such a case, it may prove worthwhile to study the possible features in a somewhat simpler context.
For this purpose we will use the two dimensional supersymmetric O(3) $\sigma$--model as a toy model. 

The SUSY $\sigma$--model shares many of the properties
of the SUSY Yang--Mills theory. Apart from being supersymmetric, they also have asymptotic freedom and instanton solutions 
in common \cite{polyakov}.
Furthermore, in this toy model there also exists an exact calculation of a condensate \cite{bohr}, that spontaneously
breaks the nonanomalous discrete chiral symmetry $Z_4 \rightarrow Z_2$
(for a comprehensive review of the SUSY $\sigma$--model, see \eg \cite{sigmarev}). 
Some precaution in the analogy at finite temperature is
needed however, since by Peierls' argument one would expect a theory in only one spatial dimension to restore a broken discrete
symmetry at {\em any} nonzero temperature. Nevertheless, our hope is that this model still could serve as a mathematical laboratory
and provide some new insights to the possible scenarios these discrete symmetries may undergo at finite temperature, \eg in the
SUSY Yang--Mills theory.

The calculation is also interesting from a more technical point of view, since we find that the quadratic fluctuations
around the instanton solution
cancel between the bosons and the fermions, at all temperatures. This is a rather surprising result, 
since the different boundary conditions
imposed on the bosonic and fermionic field at finite temperature naively seems to destroy such a ``supersymmetric'' cancellation. Although this behavior
seems to be a specific property in two dimensions (in contrast to \eg SUSY SU(2) Yang--Mills in four dimensions \cite{jimojag}),
it supports the reasoning that the SUSY theories provide simple and constrained, but still nontrivial, 
examples of doable models at finite temperature.   

The paper is organized as follows. In the next section we set up the model and recall some of the
main results at zero temperature. In section 3 we generalize the necessary ingredients to
finite temperature, and section 4 is devoted to the calculation of the finite--$T$ correlator,
with some of the details given in the appendices.
Finally, in section 5 we give our conclusions, and comment on the results.  

\vskip 1cm
\noindent{\bf 2. Supersymmetric O(3) $\sigma$-model at zero temperature. }
\renewcommand{\theequation}{2.\arabic{equation}}
\setcounter{equation}{0}
\vskip 5mm

The Euclidean action of the supersymmetric O(3) $\sigma$-model in two dimensions is defined as \cite{susyaction}
\be
S= \frac{1}{2 g^2} \int  d^2\!x\, d^2\!{\theta}\, {\varepsilon}^{\alpha \beta} 
D_{\alpha}{\Phi}_a D_{\beta}{\Phi}_a  \ , \label{action}
\ee
where $a=1,2,3$ is the internal isospin index, $g$ the coupling constant,
$D_{\alpha}$ the supercovariant derivative and $\Phi_a$ a real superfield,
\be
\Phi_a (\vec x ,\theta) = \varphi_a (\vec x) + \streck{\theta} {\Psi}_a (\vec x) 
+\half \streck{\theta} \theta F_a (\vec x) \ ,
\ee
satisfying the constraint $\sum_a {\Phi}_a (\vec x,\theta) \, {\Phi}_a (\vec x,\theta)=1$. 
This model generalizes the ordinary, non--supersymmetric $\sigma$--model.

Instead of three variables and one constraint, it is convenient to use a stereographic 
projection and trade the original fields for a complex-valued unconstrained field,
\be
\Theta = \frac{\Phi_1 +i\Phi_2}{1+\Phi_3} \ .
\ee
The dynamical component fields will then be a complex scalar $\phi$ and its fermionic counterpart $\psi$.
The original fields $\varphi_a$ and ${\Psi}_a$ transform as vectors under O(3)--rotations, and based on these
transformation properties one can find the corresponding action on $\phi$ and $\psi$. Using the three Euler angles
$0 \leq \alpha \, , \, \gamma \leq 2\pi $, $0\leq \beta \leq \pi$ and defining $\lambda = \tan (\beta /2)$, we have \cite{bohr}
\be
\phi \rightarrow e^{i\gamma} \, \frac{\phi \, e^{i\alpha} -\lambda}{1+\lambda \phi \, e^{i\alpha}}
\ \ \ \ \ \ \ \ \psi \rightarrow \frac{(1+\lambda^2)\psi \,
e^{i(\alpha +\gamma)}}{(1+\lambda \phi \, e^{i\alpha})^2} \ . \label{transformation}
\ee

In order to regulate the infrared divergencies it is assumed that the Euclidean
space is restricted to a sphere of radius $R$, and the original flat
metric then becomes the conformally flat one, $g_{\mu \nu} = \Omega_{0}^2 \delta_{\mu \nu}$, with
$\Omega_{0} = (1+(\vec{x}\, ^2 /4R^2))^{-1}$. 
At the end of the calculations the flat space limit $R \rightarrow \infty$ is 
understood to be taken.
With this modification the action (\ref{action}) becomes
\be
S &=& \frac{2}{g^2} \int d^2\!x\, \Omega_{0}^2 \, \chi^{-2} \left[ \Omega_{0}^{-2} \del_{\mu} \phi^{\ast} \, \del_{\mu} \phi
+\frac{i}{2} \Omega_{0}^{-3/2}\left( \streck{\psi} \gamma_{\mu} \del_{\mu} \Omega_{0}^{1/2} \psi -
(\del_{\mu} \Omega_{0}^{1/2} \streck{\psi}) \gamma_{\mu} \psi \right) \right. \nonumber \\
&-& \left. i\Omega_{0}^{-1}\chi^{-1} \overline{\psi}\gamma_{\mu} \psi \left( \phi^{\ast} \del_{\mu} \phi
-\phi \del_{\mu} \phi^{\ast} \right) +\frac{1}{2} ( \chi^2)^{-1}(\streck{\psi} \, \streck{\psi})(\psi \psi) \right] 
\ , \label{theaction}
\ee
where $\chi = 1+ \phi^{\ast} \phi$ and $\gamma_{\mu} = \sigma_{\mu}$, $\mu =1\, ,2$ ($\sigma$ being
the Pauli matrices). 

As is well known, the $\sigma$-model possesses nontrivial field configurations that
extremize the action, the instanton solutions \cite{polyakov}. Introducing complex coordinates 
$z=x_1+ix_2$ instead of the Euclidean coordinates $x_1$ and $x_2$, any instanton solution $\phi_{{\rm inst}}$ 
is characterized by an integer $k$, the topological charge, given by
\be
k=\frac{1}{4 \pi} \int d^2 \! x\, \frac{ \left| \del_{z} \phi_{{\rm inst}} \right|^2 - \left| \del_{\streck{z}} \phi_{{\rm inst}} \right|^2 }
{(1+ \phi^{\ast}_{{\rm inst}}
 \phi_{{\rm inst}} /4)^2} \ ,
\ee
where $\del_z = \half (\del_x -i \del_y)$.

For $k \geq 0$, the minimal value of the action is obtained for holomorphic fields, satisfying $\del_{\streck{z}} \, \phi =0$,
and to this solution corresponds $4k+2$ real-valued bosonic zero modes and $4k$ fermionic ones.
Expanding around this instanton solution, 
\be
\phi = \phi_{{\rm inst}} (z) +\frac{g}{\sqrt{2} } \phi_q  \nonumber \ , \ \ 
\psi =\psi_{\rm cl} + \frac{g}{\sqrt{2} } \psi_q  \ , 
\ee
the action (\ref{theaction}) becomes in the bilinear approximation,
\be
S= S_0 +\int d^2 x\, \Omega_{0}^2 \left[\phi_{q}^{\ast} \left(-4\Omega_{0}^{-2}\frac{\del}{\del z}\chi_{0}^{-2} 
\frac{\del}{\del \streck{z}} \right) \phi_q +2i\Omega_{0}^{-3/2} \, {\streck{\psi}}_q \left( \matrix{ 0 & \del_z \chi_{0}^{-2} 
\cr \chi_{0}^{-2} \del_{\streck{z}} & 0} \right) \Omega_{0}^{1/2} \psi_q \right] \ . \nonumber \\ \label{fluctuation}
\ee
where $S_0 = 4\pi k / g^2$  is the classical action and $\chi_0 = 1+ \phi^{\ast}_{{\rm inst}}\phi_{{\rm inst}}$.

Specializing to $k=1$, the most general instanton solution is
\be
\phi_{{\rm inst}} (z) = \frac{y}{z-z_0} +c \ ,
\ee
where $|y| = \rho$ corresponds to the size of the instanton, $z_0$ to its position and $c$,
together with the phase of $y$, to a rotation
of the instanton solution in the original internal isotopic space. These three parameters are all complex, 
giving six real collective coordinates and hence six bosonic zero modes.
There are four real-valued fermionic zero modes, 
\be
\psi_0^{(i) \alpha} (z) = \theta^{(i)} \delta^{\alpha 1} \frac{y}{(z-z_0)^i} \ , \label{tzeromodes}
\ee
with $\theta^{(i)}$ ($i=1,2$) a Grassmann parameter and $\alpha$ the spinor index. Note that the fermionic zero modes,
as they stand, are not normalized.

Now consider the correlation function,
\be
\Pi^{(n)} (\vec{x}_1,\ldots, \vec{x}_n) = \langle 0|{\rm T} \{O(\vec{x_1}) \ldots O(\vec{x}_n) \} |0\rangle \ ,
\ee
where $O=\chi^{-2} \streck{\psi} (1+\gamma_5) \psi$, $T$ stands for time ordering and $\gamma_5 = \sigma_3$.
Due to conservation of chirality, it is clear that $\Pi^{(n)}$ receives no perturbative contributions.
However, similarly to \eg QCD, there exists an axial--vector current that is conserved classically
but broken by quantum effects, by a ``diangle'' anomaly \cite{sigmarev}. The change in the axial charge $Q_5$
is given by
\be
\Delta Q_5 = 4k \ ,
\ee
implying the chiral selection rule $n=2k$ for the correlator. Hence, in a given topological sector there is 
only one correlation function that does not vanish trivially, but 
can get contributions from nonperturbative, instanton effects. 

Using the chiral selection rule, the only correlator receiving a one--instanton contribution is $\Pi^{(2)}$, and this
is the correlation function we will consider. When the action (\ref{fluctuation}) (for $k=1$) is used in the path integral,
the integrations over the bosonic zero modes are traded for the collective coordinates $c$, $y$ and $z_0$, together with the
appropriate Jacobian. Similarly there is also a Jacobian associated with
the fermionic zero modes. For correlation functions like $\Pi^{(2)}$, the entire contribution is saturated
by the zero modes in this approximation. Therefore we can integrate out the non--zero modes to get determinants of 
the bosonic and fermionic non--zero eigenvalues. The measure for the integral over the collective coordinates
is then given by
\be 
I= e^{-4\pi /g_0^2} \, d^2 x_0 \, d^2 c\, d^2 y\, d \theta^{(1)}\, d\theta^{\dagger (1)}\,
d\theta^{(2)}\, d\theta^{\dagger (2)}\, ({\rm Det'}D_B)^{-1}({\rm Det'}D_F) J \ ,
\ee
where $D_B$ and $D_F$ are the operators in the quadratic fluctuation (\ref{fluctuation}) for the
bosonic and fermionic part respectively, with the prime indicating that only the non--zero modes should be taken into account,
and $J$ is the combined fermion and boson
Jacobian for the transformation to the collective coordinates. 
The subscript on $g_0$ indicates that this is the bare coupling constant which will be renormalized.
Without the zero modes, 
the supersymmetric pairing of bosonic and fermionic degrees of freedom and 
the degeneracy of non--zero eigenvalues (that still holds in the presence of the instanton \cite{fluctcancel}) normally
imply that the boson and fermion determinants cancel. However, in the presence of an infrared regularization
this is not necessarily the case. The determinants are also ultraviolet divergent, and an UV--regularization is therefore also
needed. Now, by using
the transformation rules (\ref{transformation}) it is easily seen that
$\Pi^{(2)}$ is O(3)-invariant, and since the correlator is saturated by the zero mode solutions, whose product is
invariant under O(3) transformations, the
collective coordinate integration measure $I$ has to share this invariance as well. 
Actually the
O(3) invariance requires the contribution from the determinants and the Jacobian to depend
on the collective coordinates as (in the limit $R \rightarrow \infty$)
\be 
({\rm Det'}D_B)^{-1}({\rm Det'}D_F) J \propto \left|y \right|^{-2} \left(1+\left|c\right|^2 \right)^{-2} \ ,
\ee
and since the integration measure has to be dimensionless, one finds by dimensional arguments 
\be
I = K M^2 e^{-4\pi /g_0^2} \, d^2 x_0 \, \frac{d^2 c}{\left(1+\left|c\right|^2 \right)^{2}} \,
\frac{d^2 y}{\left|y \right|^{2}} \, d\theta^{(1)}\, d\theta^{\dagger (1)}\,
d\theta^{(2)}\, d\theta^{\dagger (2)} \ ,
\ee
with $K$ a constant and $M^2$ an UV cut-off. 
Of course, to find the correct numerical factor $K$ one has to
perform the explicit calculation.

The calculation of the correlation function then gives
\be
\Pi^{(2)} (\vec{x}_1 ,\vec{x}_2) = N \Lambda^2 \ , \label{constant}
\ee
where $N$ is a numerical factor and $\Lambda$ is the scale parameter in analogy with QCD,
\be
\Lambda^2 = M^2 e^{-4 \pi / g_0^2} \ .
\ee
This result for the correlation function
is {\em a priori} reliable only at small distances, $ \Delta x =|\vec{x}_1 -\vec{x}_2| \rightarrow 0$, 
where $\Pi^{(2)}$ is
dominated by small-size instantons, but supersymmetric arguments show that neither
does there exist any multiloop corrections \cite{exactresult}, nor can it 
ever be $x$-dependent \cite{sigmarev}.
So the result (\ref{constant}) is actually exact, for all values of $\Delta x$, 
and by the cluster decomposition this implies a double-valued vacuum condensate,
\be
\langle O \rangle \propto \pm \Lambda \ ,
\ee
leaving a discrete $Z_2$ invariance for the condensate. 

\vskip 1cm
\noindent{\bf 3.  Preliminaries at finite temperature.}
\renewcommand{\theequation}{3.\arabic{equation}}
\setcounter{equation}{0}
\vskip 5mm
In this section we generalize the ingredients necessary for a calculation of the correlation
function $\Pi^{(2)}$ at finite temperature.

At finite temperature, the Euclidean space is restricted 
to the strip $\cal M$ defined 
by ${\rm Re} (z)=x_1 \in \Re$ and $0 \leq {\rm Im} (z) =x_2 \leq \beta =1/T$, where $T$ is the temperature.
Although the temporal component of the Euclidean space 
is compact at finite temperature, the spatial part
still needs to be regulated. Moreover, any infra--red cutoff has to reduce to the $T=0$ case when
the temperature vanishes. The metric will be taken to be conformally flat,
$g_{\mu \nu} = \Omega^2 \delta_{\mu \nu}$, with
\be
\Omega = \frac{h(\beta,R^2)}{h(\beta,R^2) + (\beta^2 / \pi^2)|\sinh (\pi z / \beta) |^2 } \ ,
\ee
satisfying $\Omega (x_2 +\beta ) = \Omega (x_2)$,
and where the function $h(\beta,R^2)$ satisfies 
\be
\lim_{\beta \rightarrow \infty} h(\beta,R^2) = h(0,R^2) = 4R^2 \ , 
\ee
to ensure $\lim_{\beta \rightarrow \infty} \Omega = \Omega_0$. The Euclidean action $S$ at finite temperature is then given
by the same expression as (\ref{theaction}), with the replacements $\Omega_0 \rightarrow \Omega$ and
$\int_{-\infty}^{\infty}d^2 x\, \rightarrow \int_{0}^{\beta} dx_2 \, \int_{-\infty}^{\infty} dx_1 $, and
where bosonic (fermionic) fields are periodic (antiperiodic) under $x_2 \rightarrow x_2 +\beta$.

At finite temperature there still exists an exact instanton solution, which can be deduced from the charge $k=1$
instanton by adding an infinite string of such instantons, located at $x_2 =n T^{-1} = n\beta$ with identical sizes
and rotations,
\be
\phi_{{\rm inst}}= y \sum_{n= -\infty}^{\infty} \left(\frac{1}{z-z_0 -in\beta} \right) +c =
\frac{y \pi}{\beta} \coth \left[\pi(z-z_0)/ \beta \right]  +c \ . \label{tinstanton}
\ee
This describes a periodic $k=1$ instanton solution with the same number of collective coordinates as at $T=0$. 
Note that when ${\rm Re} (z-z_0) \rightarrow \pm \infty$ the solution becomes $\lim_{{\rm Re} (z-z_0) \rightarrow \pm \infty} 
\phi_{{\rm inst}} = c \pm y\pi / \beta$,
implying that when $T >0$ the collective coordinate $y$ is entangled with the isospin rotations parametrized by $c$.
In other words, already at this level it is clear that the rotational degrees of
freedom will be more complicated than in the $T=0$ case. This behavior should 
be contrasted with the case of Yang--Mills theory in four dimensions, where there is
no mixing of the collective coordinates at any temperature \cite{grossetal}.

From the instanton solution (\ref{tinstanton}) we find the bosonic zero modes by taking derivatives with respect
to the collective coordinates, and 
the fermionic zero modes, satisfying antiperiodic boundary conditions, are found from the zero temperature
solutions (\ref{tzeromodes}):
\be 
\psi_0^{(1) \alpha} (z) = \theta^{(1)} \delta^{\alpha 1} \sum_{n=-\infty}^{\infty} (-1)^n \frac{y}{(z-z_0-in\beta)} =
\theta^{(1)} \delta^{\alpha 1} \left(
\frac{y \pi}{\beta} \right) \frac{1}{\sinh \left[ \pi (z-z_0) / \beta \right]} \ , \label{zeromode1}
\ee
and
\be
\psi_0^{(2) \alpha} (z) = \theta^{(2)} \delta^{\alpha 1} \sum_{n=-\infty}^{\infty} (-1)^n \frac{y}{(z-z_0-in\beta)^2} =
\theta^{(2)} \delta^{\alpha 1} \left( \frac{y \pi^2}{\beta^2} \right) \frac{\cosh \left[\pi (z-z_0) / \beta \right]}
{\sinh^2 \left[ \pi (z-z_0) / \beta \right]} \ . \label{zeromode2}
\ee

At zero temperature, the Greens functions $\Pi^{(n)}$ that are not trivially vanishing but
receive instanton contributions, are determined by
the chiral selection rule. Since we are interested in the instanton corrections at finite temperature,
it is necessary to establish a corresponding selection rule at $T>0$.
Considering the correlation function at finite temperature and
denoting by $O_i = O(\vec{x}_i)$,
\be
\langle T(O_1 \cdots O_n)\rangle = 
\int {\cal D} \streck{\psi}\, {\cal D} \psi \, {\cal D} \phi^{\ast} \, 
{\cal D} \phi \, (O_1 \cdots O_n) e^{-S} \ , \label{pathintegral}
\ee
we now perform a global chiral rotation of $\psi $, 
\be
\psi (\vec{x}) \rightarrow \psi^{'}(\vec{x}) = e^{i\alpha \gamma_5}\psi (\vec{x}) \ . 
\ee
Note that any transformation of the fields has to respect the appropriate boundary conditions. For a generic chiral
parameter $\alpha (\vec{x})$, the antiperiodicity of the fermionic field requires $\alpha (\vec{x})$ to be periodic, which
is trivially satisfied in this particular case, $\alpha (\vec{x}) =\alpha = {\rm constant}$. 
The crucial step is then
to notice that the relation for the axial anomaly remains the same at any temperature; intuitively this is rather
clear, since the anomaly relation can be viewed as a short distance effect and should therefore be
independent of the influence from the medium \cite{itomull, liuni, smilga}. Using the Fujikawa method \cite{fujikawa}, we can write
\be
\langle T(O_1\cdots O_n)\rangle &=& \int {\cal D} \streck{\psi}^{'}\, {\cal D} \psi^{'} \, {\cal D} \phi \, {\cal D} \phi^{\ast} \,
(O_1^{'} \cdots O_n^{'}) e^{-S^{'}} = \nonumber \\
&=& \int {\cal D} \streck{\psi}\, {\cal D} \psi \, {\cal D} \phi \, {\cal D} \phi^{\ast} \,
\exp \left[ 2in\alpha - 4i\alpha \int d^2 x \, \tilde{k} \right] \,
 (O_1 \cdots O_n) e^{-S} \label{fujikawapathintegral} \ , \nonumber \\
\ee
where $\tilde{k}$ is the topological charge density, 
\be
k=\int d^2x \, \tilde{k} \ ,
\ee
and the integration is performed over the strip $\cal M$.
The first term in the exponential comes from the rotation of $O_1 \cdots O_n$ and the second from the change in the
measure.
Comparing (\ref{fujikawapathintegral}) with 
(\ref{pathintegral}), we get the integrated chiral Ward identity
\be
\left( 2n - 4k \right) \langle T(O_1 \cdots O_n)\rangle = 0 \ .
\ee

As in the zero temperature case, we expand the action around the instanton solution up to quadratic fluctuations.
The classical action associated with the instanton solution is not affected by the temperature, and thus
$S_0 = 4\pi /g^2$ at any temperature. We again get a Jacobian associated with the 
integrations over the bosonic collective 
coordinates and also a Jacobian from the expansion of the fermionic field in
Grassmann coefficients, belonging to the zero modes. Furthermore, the bosonic non--zero eigenvalues give a determinant
for a differential operator 
defined on the space of periodic functions, and the fermionic non--zero modes give a determinant for a 
differential operator defined on antiperiodic functions. The 
correlation function is then calculated to this order by using the integration measure derived from the semiclassical expansion and 
replacing the fields in $O_i$ by the 
bosonic instanton solution and the fermionic zero modes. At $T=0$ there is a cancellation between the bosonic and fermionic
contributions to the determinants, that can be understood as a consequence of supersymmetry. At $T\neq 0$ the different
boundary conditions for the differential operators belonging to the bosonic and fermionic sector seem to remove the a priori
reason for such a cancellation, but as we show below and in Appendix A, it still takes place. This calculation is along the
same lines as the corresponding one at zero temperature, given in \cite{bohr, fateev}. Following that calculation we first
remove the singularities in the quadratic fluctuations by
making the following rescalings:
\be
\tilde{\phi} &=& \frac{\phi_q}{\tilde{\chi}} \left(\frac{\beta^2}{\pi^2}\right) \sinh^2 \left[\pi (z-z_0) / \beta \right] \ , \label{rescales} \\
\tilde{\psi} &=& \frac{\psi_q}{\tilde{\chi}} \left(\frac{\beta^2}{\pi^2}\right) \sinh^2 \left[\pi (z-z_0) / \beta \right] \ , \nonumber
\ee
where 
\be
\tilde{\chi} &=& \chi_0 \left(\frac{\beta^2}{\pi^2}\right) \left| \sinh \left[\pi (z-z_0) / \beta \right] \right|^2 \ .
\ee
In the bilinear approximation, we now rewrite the action (\ref{fluctuation}) in terms of the rescaled variables (\ref{rescales}),
\be
S &=& S_0+ \int d^2 x\, \Omega^2 \left[\tilde{\phi}^{\ast} \left(-4\Omega^{-2} \tilde{\chi} \frac{\del}{\del z}\tilde{\chi}^{-2} 
\frac{\del}{\del \streck{z}} \tilde{\chi}  \right) \tilde{\phi} +2i\Omega^{-3/2} \streck{\tilde{\psi}} \left( 
\matrix{ 0 & \tilde{\chi} \del_z \tilde{\chi}^{-1} 
\cr \tilde{\chi}^{-1} \del_{\streck{z}} \tilde{\chi} & 0} \right) \Omega^{1/2} \tilde{\psi} \right] \nonumber \\
&=& S_0+ \int d^2 x\, \Omega^2 \left[\tilde{\phi}^{\ast} D_B \tilde{\phi} +\streck{\tilde{\psi}}D_F \tilde{\psi} \right] \ .
\ee

As shown in Appendix A, by defining
\be
f_{\mu}= \left( \matrix{1 \cr (\beta /2\pi ) \sinh (2\pi z/ \beta ) \cr (\beta^2 / \pi^2) \sinh^2 
(\pi z/ \beta ) } \right) \ , \ \ f^{'}_i 
= \left( \matrix{ \cosh (\pi z/ \beta) \cr (\beta / \pi) \sinh (\pi z/ \beta)} \right) \ , \label{ffunctions}
\ee
and 
\be
R_{\lambda \rho} = \int d^2 \! x \, \frac{\Omega^2}{\tilde{\chi}^2} f^{\ast}_{\lambda} f_{\rho} \ , \ \ 
R^{'}_{kl} = \int d^2 \! x \, \frac{\Omega}{\tilde{\chi}^2} f^{' \ast}_{k} f^{'}_{l} \ , \label{rfunctions}
\ee
the total Jacobian can be written as
\be
J= \left( \frac{1}{\left| y \right|^2} \right) \frac{{\rm Det} R}{{\rm Det} R^{'}} \ , \label{thejacobian}
\ee
where $R^{'}$ denotes the part from the fermion Jacobian and $R$ from the boson Jacobian.

As for the calculation of the determinants ${\rm Det}^{'} D_B$ and ${\rm Det}^{'} D_F$, we find, by varying with
respect to the instanton parameters (see Appendix A for details):
\be
\delta \left( \ln {\rm Det'} D_B \right) &=& \delta \left( \ln {\rm Det} R  \right) 
+\int d^2x \,\delta \left( \ln \tilde{\chi} \right) \left[ \frac{1}{\pi} \del_{\mu} \del_{\mu}\ln \tilde{\chi} -
\frac{1}{2\pi} \del_{\mu} \del_{\mu} \ln \Omega \right] \ , \label{thebosonpiece} \\
\delta \left( \ln {\rm Det'} D_F \right) &=& \delta \left( \ln {\rm Det} R^{'} \right) + \int d^2x \, \delta \left( \ln \tilde{\chi} \right) 
\left[ \frac{1}{\pi} \del_{\mu} \del_{\mu} \ln \tilde{\chi} \right] \ . \label{thefermipiece}
\ee
The second term in the square brackets of (\ref{thebosonpiece}) is calculated in the limit where the IR--regularization is removed,
$R\rightarrow \infty$;
\be
\delta \left( \frac{1}{2\pi} \int d^2 x\, \ln (\tilde{\chi}) \del_{\mu} \del_{\mu} \ln \Omega \right) =
-\delta \ln \left( (1+|c+\tilde{y}|^2)(1+|c-\tilde{y}|^2) \right) \ ,
\ee
where $\tilde{y} = y\pi /\beta$. Hence, up to some numerical factor,
\be
 J \left( {\rm Det'} D_B \right)^{-1} \left( {\rm Det'} D_F \right) = M^2  \left( \frac{1}{\left|y \right|^2} \right)
\left( \frac{1}{(1+|c+\tilde{y}|^2)(1+|c-\tilde{y}|^2)} \right) \ ,
\ee
where we have inserted the ultraviolet cut--off $M$. 

The integration measure at finite temperature can now be written as,
\be
I = \Lambda^2 \int d^2 x_0 \, \frac{d^2 c}{\left(1+\left|c+\tilde{y}\right|^2 \right) 
\left(1+\left|c-\tilde{y}\right|^2 \right)} \,
\frac{d^2 y}{\left|y \right|^{2}} \, d\theta^{(1)}\, d\theta^{\dagger (1)}\,
d\theta^{(2)}\, d\theta^{\dagger (2)} \ , \label{tempmeasure}
\ee
and this measure is invariant under O(3)--transformations, as it should.

\vskip 1cm
\noindent{\bf 4. The correlation function}
\renewcommand{\theequation}{4.\arabic{equation}}
\setcounter{equation}{0}
\vskip 5mm

In this section we perform the explicit calculation of the spatial correlation function 
\be
\Pi^{(2)} (x_1) = \left \langle \left(\frac{\streck{\psi}(x_1,0) \, (1+\gamma_5)\psi (x_1,0)}
{\chi^2 (x_1,0)}\, \frac{\streck{\psi}(0,0) \, (1+\gamma_5)\psi (0,0)}{\chi^2 (0,0)}\right) \right \rangle \ .
\ee

In the semiclassical approximation, when we replace the fields $\chi$ and $\psi$ by their classical values, we have
\be
(1+\gamma_5) \psi &=& \frac{2y\pi}{ \beta} \left( \theta^{(1)} \frac{1}{\sinh \left[ 
\pi (z-z_0) / \beta \right]} + \theta^{(2)} \left( \frac{\pi}{\beta} \right) \frac{\cosh \left[\pi (z-z_0) / \beta \right]}
{\sinh^2 \left[ \pi (z-z_0) / \beta \right]} \right) \ , \nonumber \\
\chi &=& \chi_0 = 1+ \phi_{{\rm inst}}^{\ast} \phi_{{\rm inst}} \ ,
\ee
and inserting also the integration measure (\ref{tempmeasure}) from the previous section, the correlator becomes
\be
&&\Pi^{(2)} (z) = N\Lambda^2 \left( \frac{\pi}{\beta} \right)^6 \int d^2 x_0 \, \frac{d^2 c}{\left(1+\left|c+\tilde{y}\right|^2 \right) 
\left(1+\left|c-\tilde{y}\right|^2 \right)} \,
\frac{d^2 y}{\left|y \right|^{2}} \, d\theta^{(1)}\, d\theta^{\dagger (1)}\,
d\theta^{(2)}\, d\theta^{\dagger (2)} \times \nonumber \\ && 
\frac{ \theta^{(1)}\theta^{\dagger (1)} \theta^{(2)} \theta^{\dagger (2)} \left| y \right|^4 
\left| \sinh [\pi z / \beta] \right|^2}
{\left| \sinh [\pi (z-z_0) / \beta] \right|^4 \left| \sinh [\pi z_0 / \beta] \right|^4} \left[ \frac{1}{\{1+\phi_{{\rm inst}}^{\ast}
\phi_{{\rm inst}} (z) \} \{1+\phi_{{\rm inst}}^{\ast}\phi_{{\rm inst}} (0) \} } \right]^2 \nonumber \\
&&= N\Lambda^2 \left( \frac{\pi}{\beta} \right)^6\int d^2 x_0 \, \frac{d^2 c}{\left(1+\left|c+\tilde{y}\right|^2 \right) 
\left(1+\left|c-\tilde{y}\right|^2 \right)} \,
\frac{d^2 y}{\left|y \right|^{2}} \times \nonumber \\
&& \frac{ \left| y \right|^4 \left| \sinh [\pi z / \beta] \right|^2}
{\left| \sinh [\pi (z-z_0) / \beta] \right|^4 \left| \sinh [\pi z_0 / \beta] \right|^4} \left[ \frac{1}{\{1+\phi_{{\rm inst}}^{\ast}
\phi_{{\rm inst}} (z) \} \{1+\phi_{{\rm inst}}^{\ast}\phi_{{\rm inst}} (0) \} } \right]^2 \ , \nonumber \\
\ee

where $N$ is a numerical factor. Although not written out explicitly, the instanton field $\phi_{{\rm inst}}$ also depends 
on $c$, $z_0$ and $y$.
In Appendix B we show how to perform the $c$--integration by using the Fadeev--Popov method, 
after which the correlation function becomes
\be
&&\Pi^{(2)} (z) = 
 N\pi \Lambda^2 \left( \frac{\pi}{\beta} \right)^6 \int d^2 x_0 \, \frac{d^2 y}{\left|y \right|^{2}} 
\left( \frac{\left| 1- \left| \tilde{y} \right|^2 \right | }{1+\left| \tilde{y} \right|^2} \right) \times \nonumber \\
&&\frac{ \left| y \right|^4 \left| \sinh [\pi z / \beta] \right|^2}
{\left| \sinh [\pi (z-z_0) / \beta] \right|^4 \left| \sinh [\pi z_0 / \beta] \right|^4} 
\left[ \frac{1}{\{1+\phi_{{\rm inst}}^{\ast}
\phi_{{\rm inst}} (z) \} |_{c=0} \{1+\phi_{{\rm inst}}^{\ast}\phi_{{\rm inst}} (0) \} |_{c=0} } \right]^2 
\nonumber \\ 
&&= K \Lambda^2 \pi^7 \beta^2 \left| \sinh (\pi z / \beta ) \right|^2 \int_{-\infty}^{\infty} dx_0 \, \int_{0}^{\beta} dy_0 \,
\int_0^{\infty} d\rho \, \rho^3 \left( \frac{\left| 1- (\rho^2 \pi^2 / \beta^2 )\right | }{1+ (\rho^2 \pi^2 / \beta^2 )} \right)
\times \nonumber \\
&&\left[ \left \{ \beta^2 \left| \sinh [\pi (z-z_0)/ \beta ] \right|^2 +\rho^2 \pi^2 \left| \cosh [\pi (z-z_0)/ \beta ] \right|^2 \right \}
\right. \times \nonumber \\ && \left.
\left \{ \beta^2 \left| \sinh [\pi z_0/ \beta ] \right|^2 +\rho^2 \pi^2 \left| \cosh [\pi z_0/ \beta ] \right|^2 \right \} \right]^{-2} 
\ , \label{corrstart}
\ee
where $K$ is some new constant and $\rho = |y|$. Now, writing $z_0 = x_0 +iy_0$, putting ${\rm Re} (z) =x_1$, ${\rm Im} (z) =0$
 and thus neglecting any time dependence, and defining 
\be
\tilde{x} = \frac{2\pi x_1}{\beta} \ , \ \tilde{x_0} = \frac{2\pi x_0}{\beta} \ , \ \tilde{y_0} = \frac{2\pi y_0}{\beta} 
\ , \ t=\frac{(\rho \pi / \beta)^2 -1}{(\rho \pi / \beta)^2 +1} \ \ {\rm and} \ \ u=\tilde{x_0} - \frac{\tilde{x}}{2} \ , \nonumber
\ee
we can write (\ref{corrstart}) as (dropping the tildes on the integration variables)
\be
\Pi^{(2)} (x_1) &=& \frac{K\pi \Lambda^2}{4} \sinh^2 \left( \frac{\tilde{x}}{2} \right) \int_{-\infty}^{\infty} du \, \int_0^{2\pi} 
dy_0 \, \int_{-1}^1 dt \, (1-t^2) |t| \times \nonumber \\ &&\left[ \frac{1}{\left( \cosh [u-(\tilde{x} /2)] +t\cos y_0 \right)
\left( \cosh [u+(\tilde{x} /2)] +t\cos y_0 \right) } \right]^{2} \ . \label{almostthere}
\ee
The remaining integrals are in principle straightforward, although rather nontrivial. Referring the details of the calculation to Appendix C,
we find 
\be  
\Pi^{(2)} (x_1) &=& \frac{K\pi^2 \Lambda^2}{3} \left[1+2 \pi x_1T \coth (\pi x_1T) \left( 1-2\sinh^2 (\pi x_1 T) \right) + \right.
\nonumber \\ &+& \left. 2\sinh^2 (\pi x_1T) \ln \left( 4\sinh^2 (\pi x_1T) \right) \right] \, 
\label{exactcorrelator}
\ee
where we have assumed that $x_1 >0$ for simplicity.
This result is exact in the one--instanton background and the bilinear approximation.

The correlation function is thus seen to depend on the dimensionless combination $x_1T$ in a rather complicated way,
although it is always a decreasing function of $x_1T$, as shown in Fig. 1.

\begin{figure}[h]
\epsfxsize = 12 cm
\hspace{2.7cm}
\epsfbox{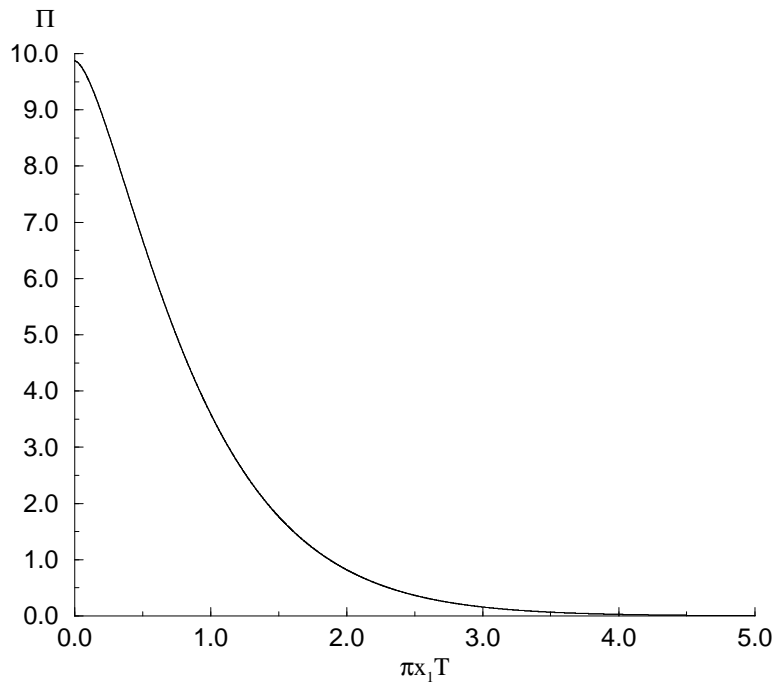}
\caption{{\em The correlator $\Pi$ in units of $K\Lambda^2$ as a function of $\pi x_1T$.}}
\end{figure}

However, for special limits it simplifies considerably. For vanishing temperature or distance,
\be
\Pi^{(2)} (x_1) \stackrel{x_1T \rightarrow 0}{\longrightarrow} K\pi^2 \Lambda^2 \propto \Lambda^2 \ . \label{smallxt}
\ee
Since the limit $x_1\ll T^{-1}$ corresponds to distances much shorter than the average separation between
the constituents of the medium, $\Delta x_{{\rm mean}} \sim T^{-1}$, there should be no temperature effects. Unsurprisingly, (\ref{smallxt}) 
is in agreement with the direct $T=0$ calculation. On the other hand, for 
asymptotically large values, $x_1T \rightarrow \infty$, the correlator falls of exponentially:
\be
 \Pi^{(2)} (x_1) \stackrel{x_1T \rightarrow \infty}{\longrightarrow} 2K\pi^2 \Lambda^2 \left( \pi x_1T \right) e^{-2\pi x_1T} \ . \label{largext}
\ee
Note that the inverse correlation length is given by twice the lowest Matsubara frequency, $\pi T$.
 
\vskip 1cm
\noindent{\bf 5. Conclusions.}
\renewcommand{\theequation}{5.\arabic{equation}}
\setcounter{equation}{0}
\vskip 5mm

We have derived, within the semiclassical expansion, an analytical expression for the instanton induced, finite temperature correlation 
function in the SUSY 2d nonlinear $\sigma$--model. For large values of the product $x_1 T$ we find that the correlator 
is exponentially decaying, and when $x_1 T \rightarrow 0$ it reduces to the well known $T=0$ calculation. We would now like to
comment on the validity of these expressions and any implications for the condensate, $\langle O \rangle$.

First of all, we have neglected all extra instanton--antiinstanton pairs and also the higher order effects beyond the
semiclassical approximation. Hence, the result for the correlator can only be accurate to the extent that the coupling is small
enough for these other effects to be neglectable. Therefore, the relevant question is what sets the scale of the running coupling.

At zero temperature, the only scale available is the distance, so in that case the semiclassical approximation
is {\em a priori} trustworthy only
at small distances. But supersymmetry guarantees that the correlation function has to be independent of the distance and thus 
makes the short distance calculation valid at all distances. However, at finite temperature the supersymmetric arguments are 
not applicable. This is rather obvious, since the correlator now
depends on $x_1$ even in the lowest approximation. 
Nonetheless, we believe the semiclassical approximation to be accurate in the high temperature regime.
The reason is that when $x_1 T \gg 1$, the correlation function is saturated by an instanton size of the order $\rho \leq T^{-1}$.
Since the relevant
scale in the correlation function should be set by the instanton size, it seems reasonable to expect the semiclassical
approximation to be valid when both $x_1 T \gg 1$ and $T \gg \Lambda$. Similarly, when $x_1T \ll 1$ the correlator should
be well described by the semiclassical approximation in the limit of vanishing distance, $x_1 \rightarrow 0$.

If the above scenario is correct, which seems plausible, the exponential decay of the correlator should be reliable at high temperatures. 
By using the cluster decomposition we then arrive at
\be
\langle O \rangle_{T \gg \Lambda} = \pm \sqrt{\lim_{x_1 \rightarrow \infty} \left[ \Pi^{(2)} (x_1) \right]_{T \gg \Lambda}} = 0 \ ,
\ee
and the discrete symmetry is restored.

Since the theory is defined in one spatial dimension, general arguments based on the free energy would imply
that the symmetry is restored at any positive temperature.
This argument is not in conflict with the above semiclassical calculation, although it is not easy to justify why the 
approximations should be qualitatively correct even at low temperatures. 

Finally, we would like to connect this model to the full, 4d SUSY Yang--Mills theory, although we must emphasize that
this is highly speculative at
the present stage. However, if the results of this toy model at finite temperature
has any generalizations to the supersymmetric Yang--Mills theory, it would
indicate that there is a high temperature phase in the 4d theory, where the discrete symmetry is restored. Such a restoration
would clearly be important in connection with the formation of domain walls \cite{domains}, and a phase transition at the temperature
of symmetry restoration can in that case not be excluded.

\vskip 1cm
\noindent
{\bf Acknowledgments}
\renewcommand{\theequation}{5.\arabic{equation}}
\setcounter{equation}{0}
\vskip .5cm
\noindent
The authors would like to thank T. H. Hansson for interesting and useful discussions. J. G. would also like to
thank the members of the theory group at the Department of Physics, Stockholm University, for their hospitality.

\vskip 1cm
\noindent{\bf Appendix A. Calculation of the Jacobian and the determinants.}
\renewcommand{\theequation}{A.\arabic{equation}}
\setcounter{equation}{0}
\vskip 5mm
In this appendix we calculate the contribution from the Jacobian and the boson and fermion determinant
at finite temperature. We will follow the $T=0$ approach \cite{bohr, fateev} and 
generalize it to the finite temperature case.

Starting with the boson Jacobian $J_B$, let $\alpha_{\mu}$ denote the collective coordinates in the following way, $\alpha_1 = z_0$, $\alpha_2 = y$, $\alpha_3 = c$.
The Jacobian is then given by
\be
J_B = {\rm Det} M^{(B)} \ , \ \ M^{(B)}_{\mu \nu} = \int d^2 \!x \, \frac{\Omega^2}{\chi_0^2} \left( \frac{ \del 
\phi_{{\rm inst}}^{\ast}}{\del \alpha_{\mu}} \right) \left( \frac{ \del \phi_{{\rm inst}}}{\del \alpha_{\nu}} \right) \ ,
\ee
where the integration is taken over the IR--regulated Euclidean strip.
By using the explicit form of the instanton solution (\ref{tinstanton}) we can write,
\be
{\rm i)}&& \ \frac{\del \phi_{{\rm inst}}}{\del \alpha_1} \left(\frac{\beta^2}{\pi^2} \right) \sinh^2 (\pi (z-z_0) / \beta ) =
y \ , \nonumber \\
{\rm ii)}&& \ \frac{\del \phi_{{\rm inst}}}{\del \alpha_2} \left(\frac{\beta^2}{\pi^2} \right) \sinh^2 (\pi (z-z_0) / \beta ) =
\nonumber \\
&&\frac{\beta}{2 \pi} \left[ -\sinh (2\pi z_0 / \beta) + \cosh (2\pi z_0/ \beta) \sinh (2\pi z/ \beta) 
-2\sinh (2\pi z_0/ \beta) \sinh^2 (\pi z/ \beta) \right] \nonumber \\
{\rm iii)}&& \ \frac{\del \phi_{{\rm inst}}}{\del \alpha_3} \left(\frac{\beta^2}{\pi^2} \right) \sinh^2 (\pi (z-z_0) / \beta ) =
\nonumber \\
&&\frac{\beta^2}{\pi^2} \left[ \sinh^2 (\pi z_0/ \beta) -\half \sinh (2\pi z_0/ \beta) \sinh(2\pi z/ \beta) 
+ \cosh(2 \pi z_0/ \beta) \sinh^2 (\pi z/ \beta) \right] \ . \nonumber \\ \label{zmrewriting}
\ee
If we now define the transposed vector
\be
f_{\mu}^T = \left[ \matrix{1 \ , & (\beta /2\pi ) \sinh (2\pi z/ \beta ) \ , & (\beta^2 / \pi^2) \sinh^2 
(\pi z/ \beta ) } \right] \ ,
\ee
and the following matrix
\be
U_{\mu \nu} = \left( \matrix{y & 0 & 0 \cr -(\beta /2\pi) \sinh (2\pi z_0 / \beta ) & \cosh (2\pi z_0/ \beta )
 & -(\pi /\beta) \sinh (2\pi z_0/ \beta ) \cr (\beta^2 / \pi^2 ) \sinh^2 (\pi z_0/ \beta ) & -(\beta / \pi)
\sinh (2\pi z_0/ \beta ) & \cosh (2\pi z_0/ \beta ) } \right) \ ,
\ee
\vskip 1mm
we can write the set of equations (\ref{zmrewriting}) in a compact form as
\be
\frac{\del \phi_{{\rm inst}}}{\del \alpha_{\mu}} \left(\frac{\beta^2}{\pi^2} \right) \sinh^2 (\pi (z-z_0) / \beta ) =
U_{\mu \nu} f_{\nu} \ .
\ee
Furthermore, by rescaling $\chi_0 = 1+\phi^{\ast}_{{\rm inst}} \phi_{{\rm inst}}$ as
\be 
\tilde{\chi} = \chi_0 \frac{\beta^2}{\pi^2} \left| \sinh [\pi (z-z_0)/ \beta ] \right|^2 \ , \label{newchi}
\ee
we find that $M_{\mu \nu}^{(B)} = U^{\dagger} R U$, where 
\be
R_{\lambda \rho} = \int d^2 \! x \, \frac{\Omega^2}{\tilde{\chi}^2} f^{\ast}_{\lambda} f_{\rho} \ . \label{boser}
\ee
The boson Jacobian is thus given by
\be
J_B = \left( {\rm Det} U \right) \left( {\rm Det} U^{\dagger} \right) \left( {\rm Det} R \right) \ . \label{jb}
\ee
Note that the definitions of $f_{\mu}$ and $U_{\mu \nu}$ reduce to the well known $T=0$ formulas in
the limit of vanishing temperature \cite{fateev}.

In the fermionic case we have to evaluate
\be
J_F = \left( {\rm Det} M^{(F)} \right)^{-1} \ ,
\ee
where
\be
M^{(F)}_{i j} = \int d^2 \! x \, \frac{\Omega}{\chi_0^2} \psi^{\dagger}_i \psi_j \ ,
\ee
and $\psi_i$ are the fermionic zero mode solutions,
\be
\psi_1 = \frac{y \pi^2}{\beta^2} \frac{\cosh (\pi (z-z_0)/ \beta )}{\sinh^2 (\pi (z-z_0)/ \beta )} \ , \ \
\psi_2 = \frac{y \pi}{\beta} \frac{1}{\sinh (\pi (z-z_0)/ \beta )} \ .
\ee
Performing the same calculations as in the bosonic case give
\be
\psi_i \left(\frac{\beta^2}{\pi^2} \right) \sinh^2 (\pi (z-z_0) / \beta ) = U'_{i j} f'_j \ ,
\ee
where
\be
f^{'}_i = \left( \matrix{ \cosh (\pi z/ \beta) \cr (\beta / \pi) \sinh (\pi z/ \beta)} \right) \ , \ 
U^{'}_{ij} = \left( \matrix{ y\cosh (\pi z_0/ \beta) & -(y\pi /\beta) \sinh (\pi z_0/ \beta) \cr
-(y\beta /\pi) \sinh (\pi z_0/ \beta) & y \cosh (\pi z_0/ \beta)} \right) \ . \nonumber \\
\ee
Now, by putting
\be
R^{'}_{kl} = \int d^2 x \, \frac{\Omega}{\tilde{\chi}^2} f^{' \ast}_{k} f^{'}_{l} \ , \label{fermir}
\ee
we get $M^{(F)} = {U}^{' \dagger} R^{'} U^{'}$, and so the fermion Jacobian becomes
\be
J_F = \left( {\rm Det} {U}^{' \dagger} \right)^{-1} \left( {\rm Det} R^{'} \right)^{-1}
\left( {\rm Det} U^{'} \right)^{-1} \ . \label{jf}
\ee
Putting the two contributions $J_B$ and $J_F$ from (\ref{jb}) and (\ref{jf}) together, we obtain the total Jacobian $J$,
$J=J_B J_F$. By explicitly taking the determinants of $U$ and $U^{'}$ we finally get the result
\be
J= \left( \frac{1}{\left| y \right|^2} \right) \frac{{\rm Det} R}{{\rm Det} R^{'}} \ . \label{finaljac}
\ee

Now consider the calculation of the determinants appearing after the gaussian integration over the 
quadratic fluctuations, taken over the non--zero eigenvalues only.
Beginning with the boson determinant, we have to evaluate
\be
\left( {\rm Det'} D_B \right)^{-1} = \exp \left[-\ln {\rm Det'} D_B \right] \ ,
\ee
where the operator $D_B$ after the rescaling is given by
\be
D_B =  -4\Omega^{-2}\tilde{\chi} \del_z \tilde{\chi}^{-2} \del_{\zbar} \tilde{\chi} \ ,
\ee
and $\tilde{\chi}$ is as defined in (\ref{newchi}). This expression for the determinant is of course formal, since
it is divergent and needs to be regularized. In order to do this we will use the proper time method,  and 
define the regularized part of the determinant as \cite{fateev}
\be
\ln {\rm Det'} D_B = \lim_{\epsilon \rightarrow 0} \left[ -\int_{\epsilon}^{\infty} \frac{d t}{t} \, \left(
{\rm Tr} e^{-tD_B} -p \right) +\alpha_1 \epsilon^{-1} -\alpha_0 \ln (\epsilon ) \right] \ ,
\ee
where $p$ is the number of zero modes, being six real--valued ones in our case. The coefficients $\alpha_1$ and $\alpha_0$ are independent of 
the instanton parameters, as will be verified in the $t\rightarrow 0$ limit, so by making a variation with respect to the parameters we get, 
\be
\delta \left( \ln {\rm Det'} D_B \right) &=& \int_0^{\infty} d t \, \left( {\rm Tr}\, \left( \delta D_B \right) e^{-tD_B} 
\right) \ .
\ee
Defining $\tilde{D}_B = -4\tilde{\chi}^{-1} \del_{\zbar} \tilde{\chi}^2 \Omega^{-2}\del_z \tilde{\chi}^{-1}$, which
satisfies $D_B \Omega^{-2} \tilde{\chi} \del_z \tilde{\chi}^{-1} = \Omega^{-2} \tilde{\chi} \del_z \tilde{\chi}^{-1} \tilde{D}_B$,
allows us to write
\be 
{\rm Tr}\, (\delta D_B ) e^{-tD_B} = 2 {\rm Tr}\, \left[ \delta (\ln \tilde{\chi} ) \left( D_B e^{-tD_B} -
\tilde{D}_B e^{-t\tilde{D}_B} \right) \right] \ ,
\ee
and hence
\be
\delta \left( \ln {\rm Det'} D_B \right) = -2 {\rm Tr} \left. \left[ \delta (\ln \tilde{\chi} ) \left( e^{-tD_B} -
e^{-t\tilde{D}_B} \right) \right] \right|^{\infty}_{t=0} \ .
\ee
Now, when $t \rightarrow \infty$ it is evident that only the zero modes contribute; there are six real valued zero
modes associated with $D_B$ and none with $\tilde{D}_B$. Thus
\be
\delta \left( \ln {\rm Det'} D_B \right) = -2 \int d^2 x\, \delta (\ln \tilde{\chi} ) P_0^B (x) +
\lim_{t \rightarrow 0} 2{\rm Tr} \left[ \delta (\ln \tilde{\chi} ) \left( e^{-tD_B} -
e^{-t\tilde{D}_B} \right) \right] \ , \label{bosedet}
\ee
where $P_0^B$ is the projection operator on the space of bosonic zero modes, $P_0^B = \sum_{\mu} 
\hat{\phi}_{\mu}^{\ast} \hat{\phi}_{\mu}$ with $\hat{\phi}_{\mu}$ an orthonormal basis.
The remaining trace can be evaluated by a heat kernel expansion of
$G_B (z,z)=\langle z| e^{-tD_B} |z\rangle$ 
and $\tilde{G}_B (z,z) = \langle z| e^{-t\tilde{D}_B} |z\rangle$ . To this end, 
consider $\lim_{z_1 \rightarrow z} G_B (z,z_1)$ and let it be represented as
\be
G_B (z,z_1) = G_B^{(0)}(z,z_1) \left[ a_0 (z,z_1) +a_1 (z,z_1) t + \ldots \right] \ , \label{expandgreen}
\ee
where $G_B^{(0)}$ is the free field solution. Since $G_B^{(0)}$ has to
satisfy periodic boundary conditions, it is obtained by summing the $T=0$ solution:
\be
G_B^{(0)}(z,z_1) = \frac{\Omega^2}{4\pi t} \sum_{n= -\infty}^{\infty} \exp \left( -\frac{\left| z-z_1 - in\beta 
\right|^2 \Omega^2}{4t} \right) \ .
\ee
The calculation of the expansion coefficients
$a_0$ and $a_1$ in (\ref{expandgreen}) is now straightforward \cite{sigmabook}. 
The value of $a_0(z,z)$ is fixed by the free field
case, $a_0 (z,z) =1$, and since we are ultimately
interested in the limit $z_1 \rightarrow z$, the result for $a_1(z,z)$
follows. We get
\be
G_B (z,z) &=& \lim_{t \rightarrow 0} \sum_{n= -\infty}^{\infty} e^{-n^2 \beta^2 \Omega^2 /4t}
\left[ \frac{\Omega^2}{4\pi t} +\frac{1}{4\pi} \del_{\mu} \del_{\mu}
\ln \tilde{\chi} \right] \nonumber \\ &=& \lim_{t \rightarrow 0}
\left[ \frac{\Omega^2}{4\pi t} +\frac{1}{4\pi} \del_{\mu} \del_{\mu} \ln \tilde{\chi} \right] \left[1+
2\sum_{n =1}^{\infty} e^{-n^2 \beta^2 \Omega^2 /4t} \right] \nonumber \\ &=& 
\frac{\Omega^2}{4\pi t} +\frac{1}{4\pi} \del_{\mu} \del_{\mu} \ln \tilde{\chi} \ .
\ee
From this equation we see that the expansion is actually independent of the boundary conditions in the
limit $t \rightarrow 0$, since the limit $t\rightarrow 0$ is equivalent to $\beta \rightarrow \infty$. This
result can be obtained by a calculation in momentum space as well \cite{Farina}. 
A similar calculation for $\tilde{D}_B$ gives
\be
\tilde{G}_B (z,z) = \frac{\Omega^2}{4\pi t} -\frac{1}{4\pi} \del_{\mu} \del_{\mu} \ln \tilde{\chi} +
\frac{1}{4\pi} \del_{\mu} \del_{\mu} \ln \Omega \ ,
\ee
and (\ref{bosedet}) is then given by
\be
\delta \left( \ln {\rm Det'} D_B \right) =-2 \int d^2 x\, \delta (\ln \tilde{\chi} ) P_0^B (x) + \int d^2x \, \delta (\ln \tilde{\chi} ) \left[
\frac{1}{\pi} \del_{\mu} \del_{\mu}\ln \tilde{\chi} -\frac{1}{2\pi} \del_{\mu} \del_{\mu} \ln \Omega \right] \ . \nonumber \\ 
\label{bosevar}
\ee
Since the variation of the determinant is unaffected by the boundary conditions, the formal
dependence on $\tilde{\chi}$ and $\Omega$
is the same as at $T=0$, but both these functions themselves are of course very different at finite
temperature. 

The remaining part in the boson determinant is to calculate the projection on the space of the zero modes.
For this purpose, we take as a non--orthonormal basis the periodic functions $g_{\mu} = \tilde{\chi}^{-1} f_{\mu}$.
The projection operator is then given by $P_0^B = \sum_{\mu , \nu} g_{\nu}^{\ast} g_{\mu} R_{\mu \nu}^{-1} \Omega^2$,
with $R_{\mu \nu}$ defined in (\ref{boser}).
Substituting this into the first term in (\ref{bosevar}),
\be
&& -2 \int d^2 x\, \delta (\ln \tilde{\chi} ) P_0^B (x) =
 -2 \int d^2 x\, \delta (\ln \tilde{\chi} )g_{\nu}^{\ast} g_{\mu} R_{\mu \nu}^{-1} \Omega^2 \nonumber \\
&=& -2 \int d^2 x\, \delta (\tilde{\chi} )\tilde{\chi}^{-3} \Omega^2 f_{\nu}^{\ast} f_{\mu} R_{\mu \nu}^{-1} =
\delta \left( \ln {\rm Det} R  \right) \ .
\ee
The boson determinant thus becomes
\be
\delta \left( \ln {\rm Det'} D_B \right) = \delta \left( \ln {\rm Det} R  \right) + \int d^2x \, \delta \left( \ln \tilde{\chi} \right)  
\left[ \frac{1}{\pi} \del_{\mu} \del_{\mu}\ln \tilde{\chi} -\frac{1}{2\pi} \del_{\mu} \del_{\mu} \ln \Omega \right]
\ . \label{bosedetfinal}
\ee

The fermionic part is calculated much in the same way; after taking the determinant over the spinor indices 
the fermion
operator with positive definite eigenvalues becomes $D_F = -4 \Omega^{-3/2}\tilde{\chi} \del_z \tilde{\chi}^{-2} \Omega^{-1} \del_{\zbar}\tilde{\chi} \Omega^{1/2}$,
and varying the regularized expression with respect to the instanton parameters we are left with
\be
\delta \left( \ln {\rm Det'} D_F \right) = -2 {\rm Tr} \left. \left[ \delta \left( \ln \tilde{\chi} \right) \left( e^{-tD_F}
-e^{-t\tilde{D}_F} \right) \right] \right|^{\infty}_{t=0} \ ,
\ee
where we have defined $\tilde{D}_F = -4 \Omega^{-3/2}\tilde{\chi}^{-1} \del_{\zbar} \tilde{\chi}^{2} \Omega^{-1} \del_{z}\tilde{\chi}^{-1} \Omega^{1/2}$,
satisfying 
\be
D_F \Omega^{-3/2} \tilde{\chi} \del_z \tilde{\chi}^{-1} \Omega^{1/2} = \Omega^{-3/2} \tilde{\chi} \del_z \tilde{\chi}^{-1} \Omega^{1/2} \tilde{D}_F \ .
\ee
As in the bosonic case there is only a contribution from the zero modes in the $t \rightarrow \infty$ limit, and when
$t \rightarrow 0$ we can make a heat kernel expansion. Expanding in powers of $t$ around the free field solution, that now
satisfies anti--periodic boundary conditions, we find 
\be
\lim_{t\rightarrow 0} \langle z| e^{-tD_F} |z \rangle &=& \lim_{t\rightarrow 0} \left[ \frac{\Omega^2}{4\pi t} +
\frac{1}{4\pi} \left( \del_{\mu}\del_{\mu} \ln \tilde{\chi} +\half \ln \Omega \right) \right] \left[ 1+ 2\sum_{n=1}^{\infty}
(-1)^n e^{-n^2 \beta^2 \Omega^2 / 4t} \right] \nonumber \\ &=& \frac{\Omega^2}{4\pi t} +
\frac{1}{4\pi} \left( \del_{\mu}\del_{\mu} \ln \tilde{\chi} +\half \ln \Omega \right) \ .
\ee
The result for $\tilde{D}_F$ is obtained by making the substitution $\tilde{\chi} \rightarrow \tilde{\chi}^{-1}$,
\be
\lim_{t\rightarrow 0} \langle x| e^{-t\tilde{D}_F} |x \rangle &=& \lim_{t\rightarrow 0} \left[ \frac{\Omega^2}{4\pi t} +
\frac{1}{4\pi} \left( -\del_{\mu}\del_{\mu} \ln \tilde{\chi} +\half \ln \Omega \right) \right] \left[ 1+ 2\sum_{n=1}^{\infty}
(-1)^n e^{-n^2 \beta^2 \Omega^2 / 4t} \right] \nonumber \\ &=& \frac{\Omega^2}{4\pi t} +
\frac{1}{4\pi} \left( -\del_{\mu}\del_{\mu} \ln \tilde{\chi} +\half \ln \Omega \right) \ .
\ee
Thus we see again that in the limit of vanishing $t$, the expansion becomes independent of the boundary conditions.
Including the projection over the zero modes as well,
\be
\delta \left( \ln {\rm Det'} D_F \right) = -2 \int d^2 x\, \delta \left( \ln \tilde{\chi} \right) P_0^F (x) +\int d^2x \, 
\delta \left( \ln \tilde{\chi} \right) \left[ \frac{1}{\pi}
\del_{\mu} \del_{\mu} \ln \tilde{\chi} \right] \ .
\ee
Taking the anti--periodic functions $g^{'}_{i}= (1/ \sqrt{\Omega}) \tilde{\chi}^{-1} f^{'}_{i}$ as a basis, 
we can write
$P_0^F = \sum_{i,j} g^{' \ast}_{j} g^{'}_i R_{ij}^{' -1} \Omega^2$, with $R^{'}_{ij}$ being the same as 
in (\ref{fermir}). Hence
\be
&&-2 \int d^2 x\, \delta \left( \ln \tilde{\chi} \right) P_0^F (x) = -2\int d^2 x\, \delta \left( \ln \tilde{\chi} \right)
g^{' \ast}_{j} g^{'}_i R_{ij}^{' -1} \Omega^2 \nonumber \\ &=& -2 \int d^2 x\, \delta (\tilde{\chi}) \tilde{\chi}^{-3} f^{' \ast}_j f^{'}_i
R_{ij}^{' -1} \Omega = \delta \left( \ln {\rm Det} R^{'} \right) \ ,
\ee
and the variation of the fermion determinant finally becomes
\be 
\delta \left( \ln {\rm Det'} D_F \right) = \delta \left( \ln {\rm Det} R^{'} \right) +\int d^2x \, \delta \left( \ln \tilde{\chi} \right)
\left[ \frac{1}{\pi} \del_{\mu} \del_{\mu} \ln \tilde{\chi} \right] \ . \label{fermidetfinal}
\ee

\pagebreak
\noindent{\bf Appendix B. Integrating out the coordinate $c$.}
\renewcommand{\theequation}{B.\arabic{equation}}
\setcounter{equation}{0}
\vskip 5mm

Here we derive the formula (\ref{corrstart}), by using the Fadeev--Popov method.
The starting point is the expression for the correlation function with all integrations over the collective
coordinates remaining,
\be
\Pi^{(2)} (z) = \int \frac{ d^2c \, d^2y \, d^2 z_0 }{\left |y \right|^2 \left(1+\left|c+\tilde{y} \right|^2
\right)  \left(1+\left|c-\tilde{y} \right|^2 \right)} f(z,c,y,z_0) \ ,
\ee
where $\tilde{y} = y \pi / \beta$ and we have defined a function
$f(z,c,y,z_0)$ that contains the scale factor $\Lambda$, the contribution from the fields in the correlation function
and some irrelevant numerical factor.

The basic idea is that by an O(3) transformation we can always make $c=0$. The integration over $c$ can thus be replaced
by an integral over O(3) parameters and a suitable Jacobian. To find the Jacobian we define
\be 
\frac{1}{F(c,y)} = \int d\Omega \, \delta(c') \ , \label{trick}
\ee
where $c'$ is the value of $c$ after an O(3)--rotation,
\be
c' = \frac{e^{i\gamma}}{2} \left( \frac{(c+\tilde{y})e^{i\alpha} - \lambda}{1+\lambda(c+\tilde{y})e^{i\alpha}} +
\frac{(c-\tilde{y})e^{i\alpha} - \lambda}{1+\lambda(c-\tilde{y})e^{i\alpha}} \right) \ ,
\ee
and $d\Omega$ is the invariant group measure, which in terms of our O(3)--parameters reads
\be
\int d\Omega = \int_0^{2 \pi} d \alpha \, \int_{-\pi}^{\pi} d \gamma \, \int_0^{\infty}  \frac{d \lambda \, \lambda}
{(1+\lambda^2)^2} \ .
\ee
This construction ensures that $F$ is rotationally invariant.
Inserting the trivial factor 
\be
F(c,y)\int d\Omega \, \delta(c') =1
\ee 
into the correlation function and performing a rotation to new
values $c \rightarrow c'$, $ \tilde{y} \rightarrow \tilde{y}'$ and $z_0 \rightarrow z_0'$
we obtain, by making use of the O(3) invariance of $f(z,c,y,z_0)$,
\be
\Pi^{(2)} (z) &=& \int d\Omega \int \frac{ d^2c' \, d^2y' \, d^2 z_0' }{\left |y' \right|^2 \left(1+\left|c'+\tilde{y}' \right|^2 \right)  \left(1+\left|c'-\tilde{y}' \right|^2 \right)}F(c',y')f(z,c',y',z_0') \delta (c') 
\nonumber \\
&=& 2\pi^2 \int \frac{d^2y \, d^2 z_0}{\left| y \right|^2 \left( 1+ \left| \tilde{y} \right|^2 \right)^2}
F(0,y) f(z,0,y,z_0) \ , \label{stillF}
\ee
($\int d\Omega = 2\pi^2$). According to the definition (\ref{trick}),
\be
\frac{1}{F(0,y)} &=& \lim_{c \rightarrow 0} \frac{1}{F(c,y)} \nonumber \\
&=& \lim_{c \rightarrow 0} \int d \alpha \, d\gamma \,
\int_{0}^{\infty} \frac{d \lambda \, \lambda}{(1+\lambda^2)^2} \delta\left[ \frac{e^{i\gamma}}{2} \left(
\frac{(c+\tilde{y})e^{i\alpha} - \lambda}{1+\lambda(c+\tilde{y})e^{i\alpha}} +
\frac{(c-\tilde{y})e^{i\alpha} - \lambda}{1+\lambda(c-\tilde{y})e^{i\alpha}} \right) \right] \nonumber \\
&=& \lim_{c \rightarrow 0} 2\pi \int \frac{d \alpha \, d \lambda \, \lambda}{(1+\lambda^2)^2} \left| 1- \lambda^2
{\tilde{y}}^2 e^{2i\alpha} \right|^2 \delta \left (c-\lambda (e^{-i\alpha} +{\tilde{y}}^2 e^{i\alpha}) \right) \ .
\label{Fcalc}
\ee
The delta function can be rewritten as,
\be
&&\delta \left (c-\lambda (e^{-i\alpha} +{\tilde{y}}^2 e^{i\alpha}) \right) =  
\frac{1}{\left|c \right| \left| e^{-i\alpha}
+{\tilde{y}}^2 e^{i\alpha} \right|} \, \delta \left(\lambda - \frac{\left|c \right|}{\left| e^{-i\alpha}
+{\tilde{y}}^2 e^{i\alpha} \right|} \right) \times \nonumber \\
&& 2 \left| \frac{{\tilde{y}}^2 e^{i\alpha} -e^{-i\alpha}}
{{\tilde{y}}^2 e^{i\alpha} +e^{-i\alpha}} + \frac{{\bar{\tilde{y}}}^2 e^{-i\alpha} -e^{i\alpha}}
{{\bar{\tilde{y}}}^2 e^{-i\alpha} +e^{i\alpha}} \right|^{-1} \, \delta \left( \alpha -\alpha (\arg (c)) \right) \ ,
\ee
and by inserting this back into (\ref{Fcalc}),
\be
\frac{1}{F(0,y)} = \frac{2\pi}{\left| \left| \tilde{y} \right|^4 -1 \right| } \ . \label{Fexpression}
\ee
With the help of (\ref{Fexpression}), we finally write (\ref{stillF}) as
\be
\Pi^{(2)}(z) = \pi \int \frac{d^2  y\, d^2 z_0}{\left| y \right|^2} \left( \frac{\left| 1- \left| \tilde{y} \right|^2
\right | }{1+\left| \tilde{y} \right|^2} \right)  f(z,0,y,z_0) \ .
\ee
Note that for $\beta \rightarrow \infty$, the effect of the $c$-integration  reduces to just an overall constant factor,
as in \cite{bohr}.

\vskip 1cm
\noindent{\bf Appendix C. Calculation of the correlation function.}
\renewcommand{\theequation}{C.\arabic{equation}}
\setcounter{equation}{0}
\vskip 5mm

In this appendix we will explicitly calculate the integrals that remain in the expression (\ref{almostthere}) for
the correlator. We start from
\be
\Pi^{(2)} (x_1) &=& \frac{K\pi \Lambda^2}{4} \sinh^2 \left( \frac{\tilde{x}}{2} \right) \int_{-\infty}^{\infty} du \, \int_0^{2\pi} 
dy_0 \, \int_{-1}^1 dt \, (1-t^2) |t| \times \nonumber \\  \nonumber \\  &\times & \left[ \frac{1}{\left( 
\cosh [u-(\tilde{x} /2)] +t\cos y_0 \right)
\left( \cosh [u+(\tilde{x} /2)] +t\cos y_0 \right) } \right]^{2} \ ,
\ee
where $\tilde{x} = 2\pi x_1 / \beta $. Next we combine the denominators by using the Feynman parameter trick and also the addition
theorem for the hyperbolic functions. The result is
\be
\Pi^{(2)} (x_1)  
&=& 3K\pi \Lambda^2\sinh^2 \left( \frac{\tilde{x} }{2} \right) \int_{0}^1 dt\, (1-t^2)t \int_0^1 d\alpha \,
(1-\alpha)\alpha \times \nonumber \\ &\times & \int_{-\infty}^{\infty} du\,  \int_0^{2\pi} dy_0  
\frac{1}{\left( \Upsilon \cosh u +t\cos y_0 \right)^4}  \nonumber \\
&=& \frac{K\pi \Lambda^2}{2}\sinh^2 \left( \frac{\tilde{x}}{2} \right) \int_{0}^1 dt\, (1-t^2)t \int_0^1 d\alpha \,
(1-\alpha)\alpha \times \nonumber \\ &\times &\int_{-\infty}^{\infty} du\,  \int_0^{2\pi} dy_0  
\int_0^{\infty} d\gamma \, \gamma^3 e^{-\gamma \left(\Upsilon \cosh u +t\cos y_0 \right)} \nonumber \\ 
&=& 2K\pi^2 \Lambda^2 \sinh^2 \left( \frac{\tilde{x}}{2} \right) \int_{0}^1 dt\, (1-t^2)t\int_0^1 d\alpha \,
(1-\alpha ) \alpha \int_0^{\infty} d\gamma \, \gamma^3 K_0 (\gamma \Upsilon) I_0 (\gamma t) \ , \nonumber \\
\ee
where 
\be
\Upsilon = \sqrt{ \cosh^2 \left( \frac{\tilde{x}}{2} \right) -(1-2\alpha )^2 \sinh^2 \left( \frac{\tilde{x}}{2} \right)} \ ,
\ee
and $K_0$ and $I_0$ are the modified Bessel functions of zero order. The integration over $\gamma$ gives a hypergeometric function,
\be
\int_0^{\infty} d\gamma \, \gamma^3 K_0 (\gamma \Upsilon) I_0 (\gamma t) = \frac{4}{\Upsilon^4} F \left( 2;2;1;t^2 / \Upsilon^2 \right) \ ,
\ee
and using the explicit form of the hypergeometric function
\be
F \left( 2;2;1;z^2 \right) = \frac{1+z^2}{\left( 1-z^2 \right)^3} \ ,
\ee
the result is,
\be
\Pi^{(2)} (x_1) = 8K\pi^2 \Lambda^2 \sinh^2 \left( \frac{\tilde{x} }{2} \right) \int_{0}^1 dt\, (1-t^2) t \int_0^1 d\alpha \,
(1-\alpha) \alpha \frac{\Upsilon^2 +t^2 }{\left( \Upsilon^2 -t^2 \right)^3 } \ .
\ee
Performing the last two integrations, and assuming $x_1 >0$ for simplicity, we finally arrive at 
\be  
\Pi^{(2)} (x_1) &=& \frac{K\pi^2 \Lambda^2}{3} \left[1+2 \pi x_1T \coth (\pi x_1T) \left( 1-2\sinh^2 (\pi x_1 T) \right) + \right.
\nonumber \\ &+& \left. 2\sinh^2 (\pi x_1T) \ln \left( 4\sinh^2 (\pi x_1T) \right) \right] 
\ee

\bibliographystyle{aip}
\bibliography{vacref}

\begin{thebibliography}{1}

\bibitem{seiwitt}
N. Seiberg and E. Witten,
\newblock Nucl. Phys. {\bf B426} (1994), 19; {\em ibid} {\bf B431} (1994), 484

\bibitem{n1seiberg}
N. Seiberg,
\newblock Nucl. Phys. {\bf B435} (1995), 129

\bibitem{Novikovetal}
V. Novikov, M. Shifman, A. Vainshtein and V. Zakharov,
\newblock Nucl. Phys. {\bf B229} (1983), 407

\bibitem{amatietal}
D. Amati, G. C. Rossi and G. Veneziano,
\newblock Nucl. Phys. {\bf B249} (1985), 1

\bibitem{domains}
G. Dvali and M. Shifman,
\newblock Phys. Lett. {\bf B396} (1997), 64; {\em erratum--ibid} {\bf B407} (1997), 452

\bibitem{linde}
A. D. Linde,
\newblock Rept. Prog. Phys. {\bf 42} (1979), 389

\bibitem{bernard}
C. Bernard et. al.,
\newblock Phys. Rev. Lett. {\bf 78} (1997), 598

\bibitem{jimojag}
J. V. Steele and J. Wirstam,
\newblock Unpublished

\bibitem{polyakov}
A. A. Belavin and A. M. Polyakov,
\newblock JETP Lett. {\bf 22} (1975), 245

\bibitem{bohr}
H. Bohr, E. Katznelson and K. S. Narain,
\newblock Nucl. Phys. {\bf B238} (1984), 407

\bibitem{sigmarev}
V. A. Novikov, M. A. Shifman, A. I. Vainshtein and V. I. Zakharov,
\newblock Phys. Rep. {\bf 116} (1984), 103

\bibitem{susyaction}
P. Di Vecchia and S. Ferrara,
\newblock Nucl. Phys. {\bf B130} (1977), 93; \\
E. Witten,
\newblock Phys. Rev. {\bf D16} (1977), 2991

\bibitem{fluctcancel}
A. D'Adda and P. Di Vecchia,
\newblock Phys. Lett. {\bf B73} (1978), 162

\bibitem{grossetal}
D. J. Gross, R. D. Pisarski and L. G. Yaffe,
\newblock Rev. of Mod. Phys. {\bf 53} (1981), 43

\bibitem{exactresult}
V. A. Novikov, M. A. Shifman, A. I. Vainshtein and V. I. Zakharov,
\newblock Phys. Lett. {\bf 139B} (1984), 389

\bibitem{fateev}
V. A. Fateev, I. V. Frolov and A. S. Schwarz,
\newblock Nucl. Phys. {\bf B154} (1979), 1

\bibitem{sigmabook}
W. J. Zakrzewski,
\newblock {\em Low--dimensional sigma models}, IOP Publishing Ltd (1989)

\bibitem{Farina}
H. Boschi--Filho, C. Farina and C. P. Natividade,
\newblock Phys. Rev. {\bf D45} (1992), 586

\bibitem{itomull}
H. Itoyama and A. H. Mueller,
\newblock Nucl. Phys. {\bf B218} (1983), 349

\bibitem{liuni}
Y.--l. Liu and G.--j. Ni,
\newblock Phys. Rev. {\bf D38} (1988), 3840

\bibitem{smilga}
A. V. Smilga,
\newblock Phys. Rev. {\bf D45} (1992), 1378

\bibitem{fujikawa}
K. Fujikawa,
\newblock Phys Rev. {\bf D21} (1980), 2848

\end{thebibliography}

\end{document}